\documentstyle[aps,prl,epsfig,multicol,times,color]{revtex} 

\newcommand{\Beginrule}{\vskip 3pt\noindent\hbox{%
\vbox{\hbox to 9cm{\hfill}}\vbox{\hrule width 9cm}} \vskip 3pt}
\newcommand{\infig}[2]{\begin{center}{\epsfig{file=#2,width=#1}}\end{center}} 
\newcommand{\StartTwoColumn}{\begin{multicols}{2}}
\newcommand{\EndTwoColumn}{\end{multicols}}

\newcommand{\Label}[1]{\label{#1}}
\newcommand{\rw}[1]{}
\def\DRAFT{%
\renewcommand{\Label}[1]{\label{##1}
{\hbox to 0cm{\textcolor{green}{\hss\em ##1\quad}}}}
\renewcommand{\rw}[1]{\vskip 10pt%
\noindent{\framebox{\textcolor{red}{New Material Needed}}}%
\par\noindent{\textcolor{red}{\em ##1}}\vskip 10pt}
\def\Input##1{\include{##1}}}

\begin{document}

\title{Evaluation of heating effects on atoms trapped in an optical trap}
\author{C.W.~Gardiner$^{1}$,  J.~Ye$^{2,3}$ , H.C.~Nagerl$^{2}$ and 
H.J.~Kimble$^{2}$}
\address{$^1$ School of Chemical and Physical Sciences, 
Victoria University, Wellington, New Zealand}
\address{$^2$ Norman Bridge Laboratory of Physics, California 
Institute of Technology 12-33, Pasadena CA 91125}
\address{$ ^3$ Present address: JILA, University of Colorado and 
\\
National 
Institute of Standards and Technology,
Boulder, Colorado 80309-0440}
 \maketitle

\begin{abstract}
We solve a stochastic master equation based on the theory of Savard 
{\em et al.} [T.A. Savard, K.M. O'Hara and J.E. Thomas, Phys. 
Rev. A{\bf 56}, R1095 (1997)] for heating arising from fluctuations 
in the trapping laser intensity. We compare with recent experiments of 
Ye {\em et. al.} [J. Ye, D.W. Vernooy and H.J. Kimble, {\em Trapping of 
single atoms in cavity QED}, 
{\small\tt quant-ph/9908007}, Phys. Rev. Lett. 
(1999), {\em in press}], and find good 
agreement with the experimental 
measurements of the distribution of trap occupancy times.  The major 
cause of trap loss arises from the broadening of the energy 
distribution of the trapped atom, rather than the mean heating rate, 
which is a very much smaller effect.
\end{abstract}
\pacs{PACS Nos. }
\begin{multicols}{2}

In a far-off resonance red-detuned trap, the effective potential of 
the trapped atom can be written
\begin{eqnarray}\Label{1}
V(x) = -{1\over 4}\alpha |{\cal E}(x)|^2
\end{eqnarray}
where $ \alpha$ is the atomic polarizability, and $ {\cal E}(x) $ is the slowly 
varying field amplitude \cite{Savard,Miller}. Following \cite{Savard}, the 
heating can be modeled using a Hamiltonian for a trapped atom of mass $ M$ of 
the form
\begin{eqnarray}\Label{2}
H = {p^2\over 2M} +{1\over 2}M\omega_{\rm tr}^2[1 + \epsilon(t)]x^2,
\end{eqnarray}
which leads to transition probabilities between trap levels of the form
\begin{eqnarray}\Label{3}
R_{n\pm 2 \leftarrow n} = {\pi \omega_{\rm tr}^2\over 16}
S_\epsilon(2\omega_{\rm tr})(n+1\pm 1)(n\pm 1).
\end{eqnarray}
In these equations, $ \epsilon(t) $ is a fluctuating quantity, whose spectrum 
is
\begin{eqnarray}\Label{4}
S_\epsilon(\omega) \equiv
{2\over \pi}\int_0^\infty d\tau\,\cos(\omega \tau)
\langle \epsilon(t)\epsilon(t+\tau)\rangle  .
\end{eqnarray}
From these transition probabilities, in follows that the time dependent 
probability $ P(n)$ that a {\em single atom} is in the $ n$th level of 
the trap under the influence of the fluctuation field satisfies the stochastic 
master equation
\begin{eqnarray}\Label{5}
\dot P(n) &= &{\Gamma_\epsilon\over 8} \bigg\{(n+2)(n+1) P(n+2)+n(n-1)P(n-2)
\nonumber \\
&& -[n(n-1) + (n+2)(n+1)]P(n)\bigg\}.
\end{eqnarray}
with the rate constant
\begin{eqnarray}\Label{6}
\Gamma_\epsilon \equiv \pi^2\nu_{\rm tr}^2S_\epsilon(2\nu_{\rm tr}).
\end{eqnarray}
\EndTwoColumn
\begin{figure}
\infig{16cm}{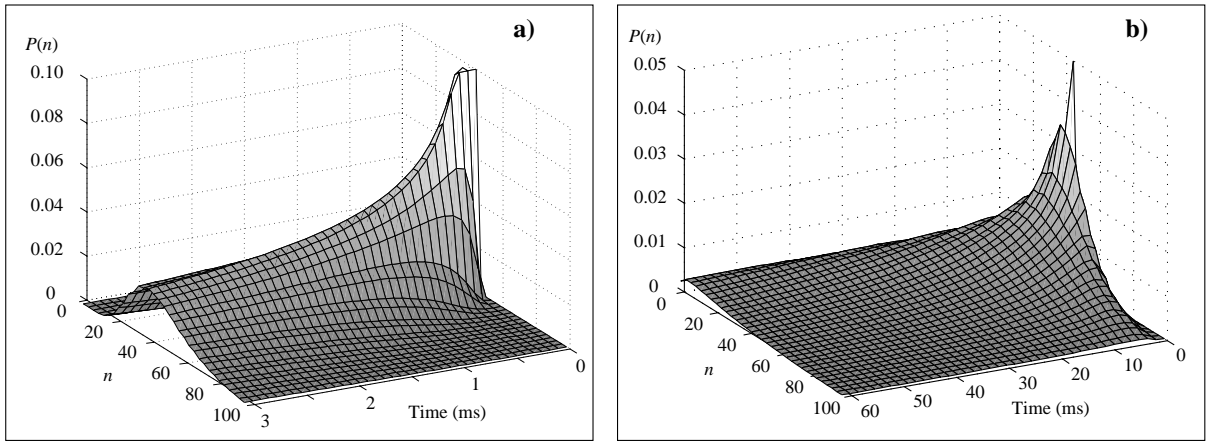}
\caption{Evolution of the probability 
\label{fig1}
distribution $ P(n)$:  a) Plotted on a short timescale, it can be seen that the 
heating spreads the initial sharp distribution in less than 2ms to cover nearly 
the full height of the trap; b) Over the full timescale of the experiment 
losses continue at a steady rate.
The heating rate used is 
$ 1/\Gamma_{\epsilon}= 1/\Gamma_{\epsilon}^{\rm axial}= 23{\rm ms}$.}
\end{figure}
\StartTwoColumn
\noindent
As shown in \cite{Savard}, this constant is equal to the {\em mean heating 
rate}, defined as the rate of 
increase of the level number (proportional to the energy) of the atom in the 
trap, i.e., 
\begin{eqnarray}\Label{7}
{d\langle n\rangle\over dt} = \Gamma_\epsilon\langle n\rangle .
\end{eqnarray} 
It should be noted, however, that this heating rate arises as the 
{\em difference} $ R_{n+2\leftarrow n} - R_{n-2\leftarrow n}$, in which the 
quadratic terms cancel.  If $ n$ is significantly different from zero---perhaps 
about 50 in \cite{Ye}---the positive and negative contributions to the heating 
rate will both  be very much larger than the heating rate itself.  Thus the 
result of the heating process will be principally to spread the distribution 
over the energy levels, superimposed on a much slower increase in the average 
energy according to (\ref{7}).  In fact, the principal time constant for the 
growth of $ \sigma$, the standard deviation of $ n$, is
$ 3\Gamma_\epsilon /2$.

The principal effect of the heating in the experiment of \cite{Ye} 
is to expel the atom from the trap, and in general this will occur not as a 
result of the increase of the average energy, but rather as a result of the 
rapid spreading of the width of the distribution, so that the upper part
spreads into untrapped levels.

The three-dimensional trap used in \cite{Ye} was sinusoidal longitudinally, and 
had a Gaussian form radially.  Approximating both of these by harmonic 
fluctuation traps, it was found by measuring the fluctuation spectrum that
\begin{eqnarray}\Label{8}
1/\Gamma_\epsilon^{\rm radial} &\approx 830{\rm ms}\\
1/\Gamma_\epsilon^{\rm axial} &\approx 23{\rm ms}.
\end{eqnarray}    
\begin{figure}
\infig{8.6cm}{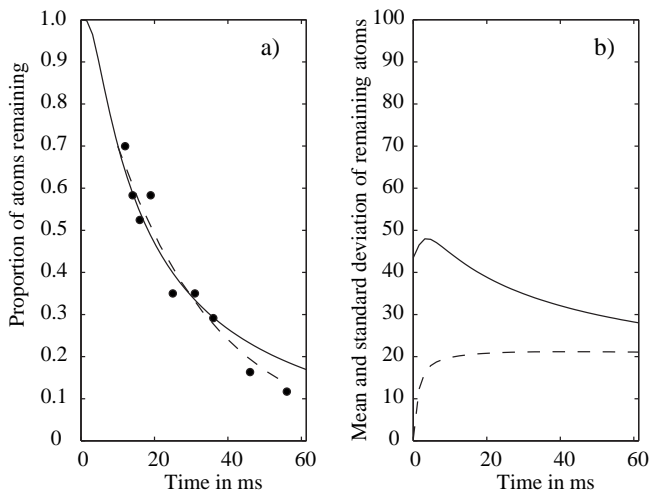}
\caption{a) Solid line: Computed probability 
for the atom to remain trapped when the initial mean excitation is the 45th 
level---heating rate as in Fig.\ref{fig1}; Points: Experimental data from 
\protect\cite{Ye}; Dashed line: 
exponential fit to data. 
b) Solid line: Mean excitation of an atom remaining in the trap; Dashed line: 
Standard deviation of the excitation.
\label{fig2} }
\end{figure}
We may safely neglect the much slower radial heating, and treat the trap as one 
dimensional.  The trap depth corresponds to some 100 levels, so we will model 
the escape process by truncating the master equation to the first 100 
levels---once the atom leaves this range it ia assumed not to return.  The 
equation is 
easy to solve.  As an initial condition, we assume the atom is evenly 
distributed between the levels $ N_0$ and $ N_0 + 1$ , with $ 0 \le N_0 < 100$.  
The results of a simulation with $ N_0=45$ are shown in Fig.1.
The very rapid spreading of the probability distribution from its initally 
sharply peaked form is very clear.  In fact very little difference results if a 
less sharply peaked initial distribution is used, even for a width of about 20 
levels.
The probability that the atom remains in the trap is plotted in Fig.~2a, and 
this fits the experimental data remarkably well. However, the result is not 
exponential, though there is a strong similarity. Points to note are
\begin{itemize}
\item From Fig.~1 and Fig.~2 it can be seen that a population around $ n=0$ is 
rapidly produced, and this decays very slowly, because the relevant transition 
probabilities are very small.  That this is not observed in practice may be the 
result of the existence of other heating mechanisms.

\item The heating rate $ \Gamma_\epsilon$ does correctly give the timescale of 
the heating process, even though the details of the heating process are not 
themselves well summarized by (\ref{7}).
\end{itemize}
To counter this heating effect one can conceive of introducing some kind of 
laser cooling.  One would expect that provided the cooling time is sufficiently 
smaller than the heating time, one should be able to ensure that the atom 
remains trapped.  We can model cooling by use of a standard master equation 
coupling to a heat bath, such as in \cite{QN}, which would give an additional 
contribution to the stochastic master equation (\ref{5})
\begin{figure}
\infig{8.6cm}{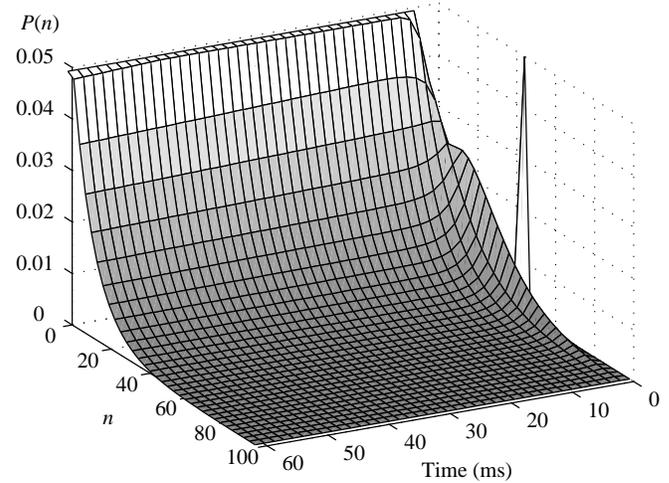}
\caption{Evolution of the probability \label{fig3}
distribution $ P(n)$ with both heating and cooling.
The heating rate used is 
$ 1/\Gamma_{\epsilon}= 1/\Gamma_{\epsilon}^{\rm axial}= 23{\rm ms}$, and the 
cooling rate is $ \Gamma_{\rm cool} = 2{\rm ms}$}
\end{figure}
\begin{eqnarray}\Label{9}
\left.\dot P(n)\right |_{\rm cool} &=& 
\Gamma_{\rm cool}\Big\{(\bar N +1)[(n+1)P(n+1)-nP(n)]
\nonumber \\&&
 +\bar N [nP(n-1) - (n+1)P(n)]\Big\}
\end{eqnarray}
In this equation  the effective temperature of the heat bath is determined by 
the mean excitation, $ \bar N$, which the bath acting by itself would produce 
in the trap, and $ \Gamma_{\rm cool} $ is the inverse cooling time. Adding this 
cooling term to the heating from (\ref{5}), we see in Fig.\ref{fig3} that the 
cooling very rapidly counteracts the heating.  However, in Fig.\ref{fig4} we 
note that even with quite strong cooling, corresponding to 
$ 1/\Gamma_{\rm cool} \approx 2{\rm ms} $, the probability of remaining in the 
trap after 60ms is only 90\%.  By solving the equations using only the cooling 
part (\ref{9}), it can be verified that most of the loss is in fact a residual 
effect of the heating.  

\begin{figure}
\infig{8.6cm}{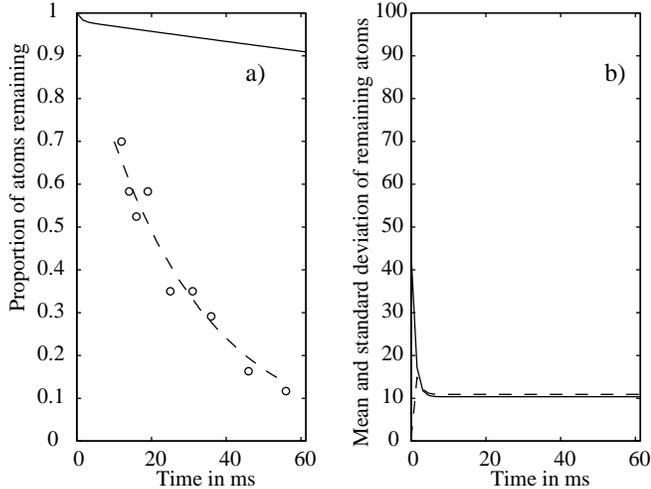}
\caption{a) Solid line: Computed probability with both heating and cooling.  
\label{fig4}
for the atom to remain trapped when the initial mean excitation is the 45th 
level---heating and cooling rates as in Fig.\ref{fig3}.; Points: Experimental 
data from \protect\cite{Ye}; Dashed line: 
exponential fit to data. 
b) Solid line: Mean excitation of an atom remaining in the trap; Dashed line: 
Standard deviation of the excitation when both heating and cooling are 
present.}
\end{figure}

However we cannot ensure better trapping simply by increasing the cooling rate, 
since the cooling has the effect of cooling to a certain residual temperature, 
and at any non-zero temperature there will always be some probability of 
escaping from the trap, even in the absence of the heating effect.  
Increasing $ 1/\Gamma_{\rm cool} $ at fixed $ \bar N $ (i.e., fixed 
temperature) is equivalent to reducing the timescale of the dynamic processes 
involved.  Once the cooling is fast enough to overwhelm the heating, any 
further increase will simply speed up the residual process of trap loss. 
The only way to get more effective confinement is then to reduce the 
temperature to which one cools.
With this model of cooling and with $ \bar N =10$, one finds that the best 
confinement is obtained with 
$ 1/\Gamma_{\rm cool} \approx 1{\rm ms} $, although this is only marginally 
better than the case of $ 1/\Gamma_{\rm cool} \approx 2{\rm ms} $ shown in the 
figures.

In conclusion one should bear in mind that the model of a truncated harmonic 
trap is very crude.  In the case considered here the noise is of the order of 
20\% of the signal, which also means that the validity of the perturnbation 
theoretic calculation used by \cite{Savard} to derive the transition 
probabilities (\ref{3}) will also be marginal at best.  However the only 
realistic alternatives to this very simple picture would involve extensive 
numerical work, such as direct simulation of a stochastic differential 
equation, or detailed computations of spectra and matrix elements for the 
appropriate potential.

\noindent
{\bf  Acknowledgements:} This work was funded by the Royal Society of New 
Zealand under the Marsden Fund cotract PVT902; by the NSF, by DARPA via the 
Quantum Information and Computing Program administered by ARO, and by the ONR.  
HCN and JY are supported by a Millikan Prize Postdoctoral Fellowship.

\end{multicols}

\begin{references}
\bibitem{Savard} T.A. Savard, K.M. O'Hara and J.E. Thomas, Phys. 
Rev. A{\bf 56}, R1095 (1997)
\bibitem{Miller} J.D. Miller, R.A. Cline and D.J. Heinzen, Phys. Rev. A{\bf 
47}, R4567 (1993)
\bibitem{Ye}  J. Ye, D.W. Vernooy and H.J. Kimble, {\em Trapping of 
single atoms in cavity QED}, {\small\tt quant-ph/9908007}, Phys. Rev. Lett. 
(1999), {\em in press}
\bibitem{QN} C.W. Gardiner, {\em Quantum Noise} (Springer, Berlin 1991); see 
Sect.6.1.
\end{references}
\end{document}